\documentclass[twocolumn,showpacs,preprintnumbers,cite,citesort,doublespace,floatfix]{revtex4}
\usepackage{graphicx}
\usepackage{dcolumn}
\usepackage{bm}

\begin{document}

\title{Entropic phase separation of linked beads}
\author{A. Zilman$^1 $, J. Kieffer$^2$, F. Molino$^2$, G. Porte$^2$ and S.A. Safran$^1$ }
\affiliation{ 1) Dept. of Materials and Interfaces, Weizmann
Institute of Science, 76100 Rehovot, Israel } \affiliation{ 2)
Groupe de Dynamique des Phasees Condensees, Universite de
Montpellier II, 34095 Montpellier Cedex 05, France }

\begin{abstract}
We study theoretically  a model system of a transient network of
microemulsion droplets connected by telechelic polymers and
explain recent experimental findings. Despite the absence of any
specific interactions between either the droplets or polymer
chains, we predict that as the number of polymers per drop is
increased, the system undergoes a first order phase separation
into a dense, highly connected phase, in equilibrium with dilute
droplets, decorated by polymer loops. The phase transition is
purely entropic and is driven by the interplay between the
translational entropy of the drops and the configurational entropy
of the polymer connections between them. Because it is dominated
by entropic effects, the phase separation mechanism of the system is extremely
robust and does not depend on the particlular physical realization
of the network. The discussed model applies as well to other systems of polymer linked particle aggregates,
such as nano-particles connected with DNA linkers.

\end{abstract}

\maketitle

\begin{figure}[!tbp]
\includegraphics[width=7cm]{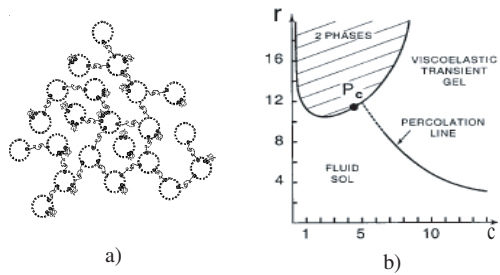}
\caption{ a)Description of the connected microemulsion. The
telechelic polymers can either link two oil droplets or loop on a
single one. b) Experimentally observed phase behavior of the
connected microemulsion as a function of the bead density $ c$ and
the mean number of stickers per droplet, $ r$.} \label{net}
\end{figure}

\emph{Introduction:} Equilibrium or transient gels are
network-forming systems, examples of which include surfactant
solutions, gels of biological
molecules or synthetic polymers \cite%
{mol,mrs,flory,talmon,degen,tanaka,strey,sack,worm,kumar}. Understanding the
phase behavior and structure of such systems is an active field of research %
\cite{rub,isaac,tanakt,big,dip,drye,miura,coniglio,mirkin}, and has a
wide range of practical applications \cite{mrs,kumar,mol,mirkin}. A
particularly elegant
experimental realization of a transient network has been reported in \cite%
{port}. The system consists of oil-in-water microemulsion droplets
(which we call either drops or beads) connected by telechelic polymers (see Fig. \ref%
{net} b)); the latter have a hydrophilic backbone with a
hydrophobic group at each chain end. Mixtures of telechelic
polymers and emulsions have a wide range of technological
applications, including paints, cosmetics and enhanced oil
recovery. Precision control of the structural and rheological
properties of materials is essential for achieving good
performance. Such control is currently achieved using the
telechelic additives, that form a transient network with
controlled rheological and structural properties \cite{port}.
Apart from its high applicative interest, the
telechelic-microemulsion mixtures serve as a model system for a
general class of transient networks. The advantage of this
particular system is that the parameters that control the
thermodynamics and structure can be easily identified and
independently controlled: the \emph{concentration of possible
nodes} (the droplets) and the \emph{connectivity of the network}
(the number of polymers per droplet). \ This is in contrast, for
example, with binary mixtures of telechelics, where one cannot
separately control the number of
nodes, formed by the associating\ chain ends. A similar system of gold nano-particles
 connected by short DNA segments, has been of a considerable interest recently \cite{mirkin, tkachenko}\\ 
 Due to the high hydrophobic energy of the stickers ($\sim 10-20 k_{\text{B}}T $ ),
the chain ends are constrained to lie within the droplets, and the
number of dangling ends outside the droplets is statistically
insignificant. Nevertheless, the polymer ends can detach from a
droplet and switch between loop (with both ends inside the same
droplet) and bridge (with\ the chain ends residing in different
droplets) configurations when the droplets are close enough.
Polymer ends can also be exchanged between the droplets during
droplet collisions. Rheological experiments show that the stress
relaxation times, which are related to the timescale for chain
rearrangement between the droplets, are of the order of one
second\cite{port}. Thus, the timescales of the observation (days)
are much larger than the time for the polymers to explore
different connectivity configurations, and the system can be
considered to be in equilibrium. \ Although the system is
\emph{athermal,} it has been found \cite{port} that for high
polymer to bead ratios, the system undergoes a first-order phase
separation. Independently, a non-thermodynamic, structural
transition was observed, where an infinite connected network is
formed, as reflected in rheological measurements. The
experimentally observed
phase diagram is shown in Fig. \ref{net} a).\\
In this Letter, we investigate theoretically the phase behavior
and structure of this experimental system. We predict that as the
average number of polymers per bead is increased, the system
undergoes a first-order phase separation into a highly connected
dense network that coexists with a dilute phase of disconnected
beads, decorated with polymer loops. This explains the experimental results of Ref. %
\cite{port} and predicts how the polymer properties control the phase
behavior. \ The predicted phase separation has a \emph{purely entropic origin};
there are no energetic interactions among the polymers or
droplets. The phase separation occurs because the loss of the
translational entropy of the droplets is overcompensated by the
high configurational entropy of the polymer connections in the
dense network. The analytical results are confirmed by Monte Carlo
simulations that also predict the phase separation.

\begin{figure}[tbp]
\includegraphics[width=7cm]{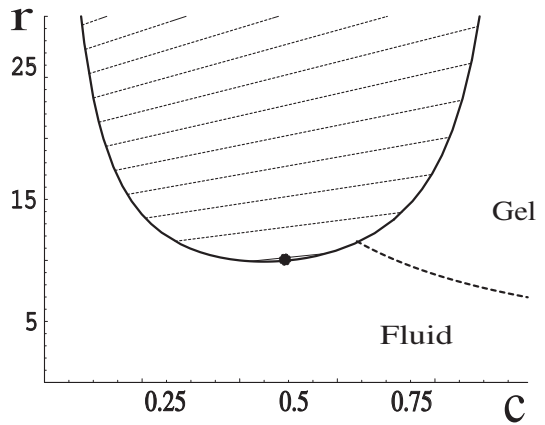}
\caption{Phase diagram of drops connected with polymers. The thick line is
the spinodal line of the phase separation for $q_{l}e^{-\protect\epsilon %
}/q=2$. Above this line the system becomes thermodynamically
unstable. The critical point is at $c=0.5$ and is shown as a black
dot. Note that the critical point is \emph{not} at the minimum of
the spinodal. The tie-lines are shown as dotted lines in the phase
separation region. Note that they are \emph{not} horizontal. The
dashed line shows the percolation threshold calculated for an
f.c.c. lattice with $q=16.$ Below the percolation line, the system
is in the fluid state, while above it a connected gel is formed. Onr should remember that the 
experimentally controlled volume fraction of the beads is equal to $c \frac{R_0^3}{(R_G+R_0)^3}$ }
\label{diag}
\end{figure}

\emph{Analytical model: }We first discuss the thermodynamic
behavior of the system. A total of $N$ beads and $N_{p}$ polymers
are distributed in space so that a polymer either connects two
beads or loops on a single bead. The total free energy of the
system, which in this athermal system amounts to the
configurational entropy, includes two contributions. The first, is
the translational entropy of the beads which we take as the
lattice-gas entropy of mixing of a dispersion of hard spheres,
$S_{0}(c)=-(c\ln c+\left( 1-c\right) \ln \left( 1-c\right) )$,
where $c$ is the volume fraction of the beads and $S_{0}(c)$ is
the entropy per site on the lattice. One should bear in mind that $ c $
is the effective volume fraction of the droplets, together with the thetered polymers.
The experimentally controlled density of pure beads is roughly $c \frac{R_0^3}{(R_G+R_0)^3} $.
The theory treating the effects of the excluded volume of the polymers, will be presented elsewhere.
The second contribution to the free energy of the system is the configurational entropy of distributing the polymers among the
beads. A polymer can either loop onto a single bead or bridge
between two droplets. For a single polymer, there are $q_{l}N $
available looped states, where $N $ is the number of the drops in
the system and $q_{l}$ is the number of positions available to a
sticker of size $a $ at the surface of a drop: $q_{l}\simeq
\frac{4\pi R_{0}^{2}}{\pi a^{2}}$. A bridging polymer connecting
two drops at a distance $R $, is stretched to a length $R$, with
an entropic cost (in
units of $k_{\text{B}}T$ and assuming ideal chains, for simplicity) $%
E_{R}\simeq \frac{3}{2}R^{2}/R_{G}^{2}$ where $R_{G}$ is the
polymer radius of gyration. From simple geometric considerations,
the average number of droplet pairs
separated by a distance $R$ is $\frac{1}{2}cN\frac{4\pi R^{2}%
}{R_{0}^{2}}g(R)\equiv q_R\,cN $ where $R_{0}$ is the radius of a
drop and $g\left( R\right) $ is the normalized density-density
correlation function of the drops. As a first step, in the mean
field approximation we take $ g(R)=1 $ for $ R>R_0 $. As we show
later, the average number of looped polymers per bead is small.
Thus, because there is no restriction on the number of polymers
connecting two given drops (except at very high polymer
concentrations, outside the range of the experiments), the
polymers are independent. The partition function of a single
polymer is $Z_1= q_l N e^{-\epsilon}+ cN\sum_R q_R\, e^{-E_{R}/T}$
where $\epsilon $ is a free energy cost (in units of  $k_B T$) of
a looped polymer, that measures the entropic cost of both ends
being confined to the same droplet.
 Taking into account the indistinguishability of the polymers,
 the configurational partition function of the total
$ N_{p}$ polymers, that determines the polymer distribution among
the beads is thus
\begin{equation}
Z_{p}=\frac{1}{N_p!}Z_1^{N_p}=\frac{\left(
qcN+q_{l}e^{-\epsilon}N\right)^{N_{p}}}{N_{p}!} \label{z}
\end{equation}%
where $q\equiv \int dR\frac{4\pi R^2}{\pi R_{0}^2}\,e^{-E_{R}/T}
$, is the effective number of droplet pairs. It is determined by
the balance of the Boltzmann factor for the stretching energy,
that strongly decreases with distance, and the number of available
positions, $ \frac{4\pi R^2}{\pi R_{0}^2} $ that increases with
distance $R $.\\
 In the thermodynamic limit, where the number of drops, polymers and
 the number of lattice sites, $V,$ all become infinite, the free energy
  per site ( in units of $k_{\text{B}}T$), $\ f\equiv -\frac{1}{V}\ln
Z_{p}-S_{0}\left( c\right) $ is
\begin{equation}
f=c\ln c+\left( 1-c\right) \ln \left( 1-c\right) +\phi \left( \ln \phi
-1\right) -\phi \ln \left( qc^{2}+cq_{l}e^{-\epsilon }\right)   \label{f}
\end{equation}
where $c=N/V$ and $\phi =N_{p}/V$ are the densities of the
droplets and of the chains  respectively. The first two terms
account for the translational entropy of the droplets and the
third term is the translational entropy of the polymer chains (the
excluded volume of the polymers is neglected at this stage). The
last term is the effective, entropically-induced attraction
between the polymers and droplets; this term causes the free
energy to decrease when the concentration of the beads or
polymers/beads ratio is increased. We stress that there are no
specific interactions between the droplets and/or the polymers,
and the resulting thermodynamic attraction is of purely entropic
origin. The system is thermodynamically stable if the free energy
$f\left( \phi ,c\right) $ (eq. (\ref{f})is a convex function of
both $\phi $ and $c $. Formally, this means that the second
variation of the free energy, $\delta ^{2}F=F_{\phi \phi }\delta
\phi ^{2}+F_{cc}\delta c^{2}+2F_{c\phi }\delta \phi \delta c$ is
positive so that the stability matrix $\hat{S}\equiv \left[
\begin{array}{cc}
F_{\phi \phi } & F_{c\phi } \\
F_{\phi c} & F_{cc}%
\end{array}%
\right] $ is positive definite i.e. possessing two positive eigenvalues.
Because $F_{\phi \phi }=\frac{1}{\phi }$ is always positive, it is
sufficient for the determinant Det $\hat{S}$ to be positive, in order to
guarantee convexity. Using Eq.  $\left( \ref{f}\right) $ for $f\left( \phi
,c\right) $, one finds Det $\hat{S}=-2\frac{1}{c\left( c+\frac{q_{l}}{q}%
e^{-\epsilon }\right) }+\frac{1}{\phi c(1-c)}$ and the system is
thermodynamically stable only if

\begin{equation}
2\phi /c<\frac{c+\frac{q_{l}}{q}e^{-\epsilon }}{c(1-c)}  \label{spinodal}
\end{equation}

where $2\phi /c\equiv r$ is the average number of stickers
(polymer chain ends) per droplet. \ This stability condition
defines the spinodal line in the $\left( c ,r\right) $ plane as
shown in Fig. \ref{diag}. Note that because the destabilizing
effect comes from the cross-terms, $F_{c\phi }$, the tie-lines are
$not$ horizontal, but connect points with different values of $r$.
If $r>\frac{c+\frac{q_{l}}{q}e^{-\epsilon }}{c(1-c)}$, the system
is thermodynamically unstable and phase separates into a system of
dense droplets that are highly connected by polymers, that
coexists with a dilute system of almost disconnected droplets,
decorated with polymer loops. The latter observation stems from
the fact that the average fraction of the looped polymers
$\bar{\lambda}$, is given by:
\begin{equation}
\bar{\lambda}=-\frac{\partial \ln(Z_1)}{\partial \epsilon}=\frac{\frac{q_{l}}{q}e^{-\epsilon }}{\frac{%
q_{l}}{q}e^{-\epsilon }+c}  \label{loops}
\end{equation}%
as follows from eq.(\ref{z}). Thus, the fraction of looped
polymers $\bar{\lambda}$ tends to unity in the dilute phase, where
$c\ll \frac{q_{l}}{q}e^{-\epsilon }$. \ From the equality of
the polymer chemical potential, $\mu _{\phi }=\ln \frac{\phi }{%
qc^{2}+q_{l}e^{-\epsilon }c}$ in the coexisting phases, it follows that the
phases that coexist lie along the lines $r=m(c+\frac{q_{l}}{q}e^{-\epsilon })
$ in the $\left( c,r\right) $ plane; where $m$ is a constant determined by $%
\mu _{\phi }$, $m=e^{\mu _{\phi }}/q$. For any given $m>4$, this
line intersects the spinodal (eq. \ref{spinodal}) at two points,
as shown in
Fig. 2.  For $m=4$ there is only one solution $\left( \frac{1}{2},2\left( 1+%
\frac{q_{l}}{2q}e^{-\epsilon }\right) \right) $ and the tie line
is tangent to the spinodal; this determines the location of the
critical point which is in our mean field model is given by
$c=0.5$ $\emph{independent}$ of the value of the control parameter
$\frac{q_{l}}{q}e^{-\epsilon }$, and the critical polymer to bead
ratio, $r_{c}=2\left( 1+\frac{q_{l}}{2q}e^{-\epsilon }\right) $.
For $m<4$ there is no phase coexistence. The predicted phase diagram of the system 
is shown in Fig. 2. Comparing the phase diagram of Fig. 2 with the 
experimentally observed one of Ref. \cite{port}

We now estimate the \ parameter $\epsilon $ that reflects the
reduction in the entropy of a polymer due to the constraint that
both ends reside in the same droplet. In a simple approximation,
the number of configurations available to a polymer with radius
$R_{G}$, with both ends constrained to a volume $v$,\ that is
small relative to the total volume
available, is proportional to $\frac{v}{R_{G}^{3}}\simeq \frac{R_{0}^{2}l}{%
R_{G}^{3}}$ where $\ l$ is the length of the hydrophobic sticker. Therefore $%
e^{-\epsilon }\simeq $ $\frac{R_{0}^{2}l}{R_{G}^{3}}$ for $R_{0}<R_{G}$ and
saturates to unity for $R_{0}>R_{G}$. The parameters $q_{l}$, $q$ and $%
\epsilon $ (which is entropic in origin), combine to yield the
single control parameter of the system: $q_{l}e^{-\epsilon }/q$.
Using typical values of the droplet radius, polymers radius of
gyration and the size of the hydrophobic stickers from
Ref.\cite{port}, we estimate the control parameter $q_l
e^{-\epsilon}/q\sim 2-3 $.\\
 It is important to realize that the
predicted phase separation \emph{cannot} be attributed to the
direct attraction between the beads induced by the connecting
polymers;
the total polymer stretching energy site is $E\simeq \frac{1}{2}\phi \frac{%
\int E_{R}e^{-E_{R}/T}q_{_{R}}\,dR}{\int
e^{-E_{R}/T}q_{_{R}}\,dR}$ and is irrelevant for the phase
separation because it is linear in \ polymer density $\phi $.

Due to its entropic nature, the phase separation is extremely
robust and is independent of the detailed assumptions about the
polymer properties. \ It persists even  in the limit of very short
and rigid chains, that  connect only nearest neighbor drops,
complementary to the case of flexible chains discussed above.
Neglecting the single bead loops (which can be shown to be
unimportant in this limit), the total number of ways to distribute
$N_{p}$ indistinguishable polymers among $P\equiv \frac{1}{2}qcN$
nearest neighbor
pairs is $\frac{\left( N_{p}+P-1\right)! }{N_{p}!\left( P-1\right) !}$ \cite%
{lanlif} and the total free energy per unit volume is $f\simeq -S_{0}\left(
c\right) +\frac{1}{2}qc^{2}\left( \ln (\frac{1}{2}qc^{2})-1\right) -\frac{1}{%
2}qc^{2}\ln \phi $ when $r=2\phi /c\gg 1$. This free energy also
exhibits a phase separation between a dense, highly connected
phase in equilibrium with a dilute solution of beads.

The mean field picture can be improved by using an \emph{exact}
mapping of the droplet-polymer system to a lattice gas model.
Denoting the chemical potential of the beads as $\mu_b $ and that
of the polymers as $\mu _{p}$, the grand canonical partition
function of the polymer-microemulsion system\ (with no loops) can
be written as
\begin{eqnarray}
\Theta  &=&\sum_{N,P}\Omega \left( N,P\right) \sum_{N_{p}=0}^{\infty }\frac{%
\left( N_p+P-1\right)! }{N_p!\left( P-1\right) !}e^{\mu_b N+\mu
_{p}N_{p}}
\nonumber \\
&=&\sum_{N,P}\Omega \left( N,P\right) e^{\mu_b N}\left( 1-e^{\mu
_{p}}\right) ^{-P}  \label{theta}
\end{eqnarray}%
here $\Omega (N,P)$ is the number of ways to arrange $N$ drops so
that $P$ nearest neighbor pairs of drops exist in \emph{a given}
realization of the grand canonical ensemble ($P$ is equal to$\
\frac{1}{2}cN$ only in the mean field approximation).  Eq.
(\ref{theta}) $exactly$ corresponds to the grand canonical
partition function of a lattice-gas with a Hamiltonian $H=\ln
(1-e^{\mu _{p}})\sum_{<i,j>}\sigma _{i}\sigma _{j}+\mu_b
\sum_{i}\sigma _{i}$. Thus, the polymer chemical potential $\mu
_{p}$, related to the number of the polymers in the system, plays
the role of the interaction parameter that controls the phase
separation. Note that this mapping can be extended off-lattice
using similar, although more complicated, arguments. The very general nauture of this 
mapping pertains to any system of linked beads, and the results described in thsi Letter
apply to many different experimental realizations, including DNA nanoparticles assemblies \cite{mirkin} 
and antigen-antibodies immunoessays \cite{mol}\\
\emph{Percolation:} The transition from a fluid-like state to an
elastic gel is related to the formation of an infinite network of
droplets connected by polymers. This is described by site-bond
percolation models that have been applied to study gelation
\cite{coniglio}. The vertices of a lattice are occupied with
probability $p_{s}$, and
a bond can form between two, occupied, nearest neighbors with probability $%
p_{b}$. Percolation is said to occur when an infinite cluster of occupied
sites, connected by the bonds, is formed. In our system, the
site-occupation probability $p_{s}$ can be identified with the droplet concentration, $c$. The "bond
occupation" probability can be identified as the probability
that at least one polymer connects a given pair of beads. For a given
numbers of chain ends per droplet, $r$, the average number of bonding
polymers per available nearest neighbor pair of drops is $r_{b}=\frac{r(1-%
\bar{\lambda})}{qc}$ where $\bar{\lambda}$ is the average fraction of looped
polymers, Eq.$\left( \ref{loops}\right) $. To a good approximation, the
probability $p_{n}$ that a given pair of nearest neighbor droplets is
connected by $n$ polymers, obeys the Poisson distribution, $p_{n}=\frac{%
r_{b}^{n}}{n!}e^{-r_{b}}$ . Thus, bond probability is $p_{b}$ $%
=\sum_{n=1}^{\infty }p_{n}=1-e^{-r_{b}}$. Previous studies of
percolation have shown \cite{holtz} that the percolation line in
the $\left( p_{b},p_{s}\right) $ plane follows a power law
$p_{b}=\left( p_{s}\right) ^{\alpha }p_{b}^{c}$ where $\alpha
=-\frac{\ln p_{b}^{c}}{\ln p_{s}^{c}}$ and $p_{b}^{c}$ and
$p_{s}^{c}$ are the percolation thresholds for independently
calculated bond and site percolation on the given lattice.
Although  lattice models for percolation cannot be applied in a
quantitative manner to the continuum, polymer-droplet system  they
provide a qualitative indication of the gelation transition as
shown in Fig. 2. \\
\emph{Simulations:}To independently verify the
predictions of the analytical model, we have performed the Monte
Carlo simulations in the grand-canonical ensemble of both beads
and polymers, with chemical potentials $\mu _{b}$ and $\mu _{p}$,
respectively. The simulated phase behavior also exhibits a first
order transition between a dense connected network and a dilute
phase. The detailed results of the simulations will be presented
elsewhere \cite{later}.\\
 \ The predicted phase diagram, summarized in Fig. 2,
reproduces the experimentally observed phase behavior of the model
transient network of Ref.\cite{port}, shown in Fig. 1 a). The
spatial phase separation described in this Letter originates from
a purely entropic effect, observed in other physical systems as
well \cite{onsager,colcrys} Namely, the loss of entropy due to
spatial inhomogeneity and formation of the dense phase, is
overcompensated by the increase in the polymer configurations. The
simplicity of the experimental realization which allows for easy
control over the system parameters, the described system can be
extended to cover a wide range of equilibrium and non-equilibrium
networks, in particular to examine the role of entropic effects in
the formation of spatially inhomogeneous structures in
non-equilibrium networks\cite{barabasi,havlin}\\
The authors thank D. Lukatsky for a critical reading of the
manuscript and N. Gov, J.F. Joanny, S. Havlin  and A. Rotem for
helpful comments. The support from GIF foundation and the PRF
fund, administered by the American Chemical Society, is gratefully
acknowledged.


\begin{thebibliography}{20}

\bibitem{mol}  M. Albers $et$ $al.$, {\it Molecular Biology of the Cell},
Garland Publishing, New York, 1994.

\bibitem{mrs}  {\it Mat. Res. Soc. proceedings, }Fall Meeting 2000,
symposium OO, MRS, Pennsylvania, 2001, {\it and references therein}.

\bibitem{flory} P. J. Flory, Principles of Polymer Chemistry, Cornell
University Press, 1981.

\bibitem{degen} P.G. de Gennes, Scaling concepts in polymer
Physics, Cornell University Press, 1979

\bibitem{talmon} A. Berheim-Groswasser $et$ $al$., Langmuir, 15: 5448
(1999); A. Berheim-Groswasser $et$ $al$., Langmuir, 16, 4131

\bibitem{sack} M. Tempel, G. Isenberg, E. Sackmann, Phys. Rev. E, 54, 1802
(1996).

\bibitem{mirkin} T. A. Taton. C. A. Mirkin, J. L. Letsinger, Science, 289, 1757 (2000)

\bibitem{tkachenko} A. V. Tkachenko, Phys. Rev. Lett., 89 (14), 148303 (2002)

\bibitem{worm} A. Khatory \textit{et. al.}, Langmuir, 9, 933 (1993); E. Buhler, Munch J. P.,
Candau J. S., J. Phys II, 5, 765 (1995)

\bibitem{kumar}  S.K. Kumar, A. Z. Panagiotopoulos, Phys. Rev. Lett., 82,
5060 (1999) {\it and references therein}

\bibitem{strey}  R. Strey \textit{et al.}, J. Chem. Phys, 105, 1175 (1996); F.
Lichterfeld $et$ $al$, J. Phys. Chem., 90, 5762 (1986){\it \ and references
therein}

\bibitem{tanaka} T. Tanaka \textit{et al.}, Phys. Rev. Lett, 42, 1556 (1979)

\bibitem{miura} T. Miura, H. Okumoto, H. Ichijo, Phys. Rev. E, 54, 6596 (1996); A.
Moussaid {\em et.al.}, J. Phys II, 1, 637 (1991) {\it and references therein}

\bibitem{isaac} T.C. Lubensky, J. Isaacson, Phys. Rev. Lett. 41, 829
(1978); T.C. Lubensky, J. Isaacson, Phys. Rev. A 20, 2130-2146 (1979)

\bibitem{tanakt} F. Tanaka, Physica A, 257, 245 (1998), F. Tanaka,
Macromolecules, 31, 384 (1998)

\bibitem{coniglio} A. Coniglio, H. E. Stanley, W. Klein, Phys. Rev. Lett.,
42, 518 (1979); A. Coniglio, H.E. Stanley, W. Klein, Phys. Rev. B
25, 6805 (1982)

\bibitem{rub} A.N. Semenov, M. Rubinstein, Macromolecules, 31, 1373 (1988)

\bibitem{drye} T. Drye, M.E. Cates, J. Chem. Phys. 96 (2), 1367 (1992); T.
Tlusty T, S.A. Safran, R. Strey, Phys. Rev. Lett., 84, 1244 (2000)

\bibitem{dip} T. Tlusty, S.A. Safran, Science, 290, 1328 (2000)

\bibitem{big} A. G. Zilman and S. A. Safran, Phys. Rev. E, 66, 051107(2002)

\bibitem{port} E. Michel {\em et. al.,}J. Rheol. 45, 1465 (2001); M.
Filali {\em et. al., }J. Chem. Phys. B, 105, 10528 (2001)

\bibitem{colcrys} W. B. Russel, D. A. Saville, W. R. Schowalter,
Colloidal Dispersions, Cambridge University Press, 1989

\bibitem{onsager} L. Onsager, Ann. N. Y. Acad. Sci., 51, 627 (1949)

\bibitem{holtz} R. L. Holtz, J. Phys, 21, 1303 (1988) {\em and references
therein}

\bibitem{barabasi} R. Albert, A. L. Barabasi, Rev. Mod. Phys. Vol.
74, 47, Jan 2002 and references therein

\bibitem{havlin} R. Cohen, D. Ben-Avraham, S. Havlin, Phys. Rev. E,
66, 036113 (2002) and references therein

\bibitem{lanlif} L. Landau and E. Lifshits, Statistical
Mechanics, vol. I, Pergamon Press, New York, 1980

\bibitem{later} A. Zilman, J. Kieffer, F. Molino, G. Porte, S.
Safran, {\em to be published}

\end{thebibliography}
\end{document}